\documentclass[conference]{IEEEtran}
\usepackage{cite}
\usepackage{amsmath,amssymb,amsfonts}
\usepackage{algorithmic}
\usepackage{graphicx}
\usepackage{textcomp}
\usepackage{xcolor}
\usepackage{xspace}
\usepackage{fontawesome}
\usepackage{multirow}
\usepackage{rotating}
\usepackage{tikz}
\usepackage{soul}
\usepackage{makecell}
\usepackage{url}
\usepackage{enumitem}
\usepackage{booktabs}
\usepackage{fontawesome}
\usepackage{pifont}
\usepackage{amssymb} 
\usepackage[colorlinks=true, allcolors=black]{hyperref}
\usepackage{tcolorbox}
\usepackage[caption=false]{subfig}

\newcommand{\ie}{\emph{i.e.,}\xspace}
\newcommand{\eg}{\emph{e.g.,}\xspace}
\newcommand{\etc}{etc.\xspace}
\newcommand{\etal}{\emph{et~al.}\xspace}
\newcommand{\secref}[1]{Section~\ref{#1}\xspace}

\newcommand{\figref}[1]{Fig.~\ref{#1}\xspace}

\newcommand{\tabref}[1]{Table~\ref{#1}\xspace}

\newboolean{showcomments}
\setboolean{showcomments}{true}

\ifthenelse{\boolean{showcomments}}
  {\newcommand{\nb}[2]{
    \fbox{\bfseries\sffamily\scriptsize#1}
    {\sf\small$\blacktriangleright$\textit{#2}$\blacktriangleleft$}
   }
  }
  {\newcommand{\nb}[2]{}
  }

\def\BibTeX{{\rm B\kern-.05em{\sc i\kern-.025em b}\kern-.08em
    T\kern-.1667em\lower.7ex\hbox{E}\kern-.125emX}}

\begin{document}

\title{Quality In, Quality Out: Investigating Training Data's Role in AI Code Generation\vspace{-0.1cm}}

\author{
\IEEEauthorblockN{Cristina Improta\IEEEauthorrefmark{1}, Rosalia Tufano\IEEEauthorrefmark{2}, Pietro Liguori\IEEEauthorrefmark{1}, Domenico Cotroneo\IEEEauthorrefmark{1}, Gabriele Bavota\IEEEauthorrefmark{2}}
\IEEEauthorblockA{\IEEEauthorrefmark{1}\textit{University of Naples Federico II, Naples, Italy}}

\IEEEauthorblockA{\IEEEauthorrefmark{2}
\textit{Università della Svizzera italiana (USI), Switzerland}\vspace{-0.2cm}}}

\maketitle

\begin{abstract}

Deep Learning (DL)-based code generators have seen significant advancements in recent years. Tools such as GitHub Copilot are used by thousands of developers with the main promise of a boost in productivity. However, researchers have recently questioned their impact on code quality showing, for example, that code generated by DL-based tools may be affected by security vulnerabilities. Since DL models are trained on large code corpora, one may conjecture that low-quality code they output is the result of low-quality code they have seen during training. However, there is very little empirical evidence documenting this phenomenon. Indeed, most of previous work look at the frequency with which commercial code generators (\eg Copilot, ChatGPT) recommend low-quality code without the possibility of relating this to their (publicly unavailable) training set. In this paper, we investigate the extent to which low-quality code instances seen during training affect the quality of the code generated at inference time. We start by fine-tuning a pre-trained DL model on a large-scale dataset ($>$4.4M functions) being representative of those usually adopted in the training of code generators. We show that 4.98\% of  functions in this dataset exhibit one or more quality issues related to security, maintainability, coding practices, \etc We use the fine-tuned model to generate 551k Python functions, showing that 5.85\% of them are affected by at least one quality issue. We then remove from the training set the low-quality functions, and use the cleaned dataset to fine-tune a second model which has been used to generate the same 551k Python functions. We show that the model trained on the cleaned dataset exhibits similar performance in terms of functional correctness as compared to the original model (\ie the one trained on the whole dataset) while, however, generating a statistically significant lower number of low-quality functions (2.16\%). Our study empirically documents the importance of high-quality training data for code generators.
\end{abstract}


\section{Introduction} \label{sec:intro}




AI-assisted software development is nowadays a reality \cite{Ernst:ieee2022}. In just a few years, we moved from code completion tools able to recommend the next few tokens the developer is likely to write \cite{Perelman:pldi2012,GveroKKP13,kyaw2018proposal,wang2020towards}, to Deep Learning (DL)-based solutions able to support code generation \cite{githubcopilot,codewhisperer,chatgpt}, namely the task of automatically implementing code described in natural language. Several studies demonstrated the benefits brought by the usage of AI-based coding assistants \cite{Arghavan:jss2023,Chatterjee2024TheIO}, for example in terms of increased perceived and actual productivity \cite{Ziegler:acm2024}. However, recent work also looked at the downsides of relying on these tools, especially when it comes to the quality of the code they generate \cite{pearce2022asleep, fu2023security, hajipour2024codelmsec, taeb2024assessing, majdinasab2023assessing, yeticstiren2023evaluating, Siddiq:scam2022, Cotroneo:icpc2024}.
Most of these works focus on security (\ie the presence of vulnerabilities in the generated code) and look at the code generated by commercial solutions such as Copilot \cite{githubcopilot}, CodeWhisperer \cite{codewhisperer}, and ChatGPT \cite{chatgpt}. 

Such a design choice, while resulting in studying widely adopted tools, does not allow to investigate whether the low-quality solutions are the results of (low-quality) coding patterns spread in the training set, which is usually not publicly available. 
Some researchers tried to overcome this problem by fine-tuning models on datasets affected by quality issues to then study how this impacts the quality of the code at inference time. For example, Cotroneo \etal \cite{Cotroneo:icpc2024} fine-tuned DL models for the task of code generation on datasets affected by a different percentage of vulnerable functions, showing that increasing the number of vulnerable training instances results in a consequent increase of vulnerabilities in the generated code. However, their work mainly aims at simulating a data poisoning attack in which the attacker introduces vulnerable instances in the training set: the authors rely on small (artificially built) training sets featuring a few hundreds instances ($<$700 functions), thus not being representative of the large-scale datasets used in practice for the training of code generators. Siddiq \etal \cite{Siddiq:scam2022}, instead, focus on more realistic training sets (up to 250k functions) and expand the analysis to a broader notion of code quality including, besides vulnerabilities, also code smells \cite{Fowler:refactoring}. However, they do not look at how these quality issues propagate into the code generated by the model, \ie what happens when training a model on code featuring quality issues \emph{vs} the same model trained on a ``cleaned dataset'', not including low-quality code. Basically, there is no evidence in the literature documenting whether actively curating the training data by removing low-quality instances is the right investment to perform in order to boost the quality of code generated at inference time.

We present a large-scale study explicitly investigating \textit{(i)} the extent to which code quality issues affecting the training set of DL-based code assistants result in the generation of low-quality code at inference time; \textit{(ii)} whether excluding these quality issues from the training data improves the quality of generated code without affecting its correctness.
We download as training dataset the subset of Python code in The Stack \cite{thestack}, a publicly available dataset featuring 6TB of code written in 30  languages. After some filtering steps, we ended up with a training dataset featuring 4.4M instances (\ie pairs $\langle$description, code$\rangle$ representing Python functions with their textual description). We then use the Semgrep static analyzer \cite{semgrep} to identify low-quality functions in such a dataset, showing that 4.98\% of its functions ($>$220k) are affected by at least one quality issue (\eg maintainability issues), with an additional $\sim$1\% ($\sim$35k) being syntactically incorrect. 

This is a first outcome of our study: \emph{large-scale code datasets feature a non-negligible percentage of low-quality and syntactically wrong code instances}. 

Then, to assess the impact of these quality issues on the generated code, we perform a two-step evaluation. First, we use the built training dataset to fine-tune DeepSeek-Coder \cite{guo2024deepseek} for the task of code generation. 

DeepSeek-Coder is a Transformer-based~\cite{vaswani2017attention} model that has been pre-trained on a code corpus featuring 2 trillion tokens, including Python code. Once fine-tuned, we asked DeepSeek-Coder to generate 551k Python functions and ran Semgrep on them looking for low-quality code. We found that 5.85\% of the generated functions are affected by quality issues, with an additional 7.13\% featuring syntactic errors. Next, to assess whether there is a relationship between the low-quality code in the training set and the quality issues generated at inference time, we remove from the training set all low-quality and syntactically wrong functions that Semgrep is able to find, fine-tuning DeepSeek-Coder again on the new, cleaned, data. 

We show that such a model, when asked to generate the same 551k Python functions, produces a statistically significant lower number of low-quality (2.16\%) and syntactically wrong (6.68\%) functions. This leads to the second and main outcome of our study: \emph{low-quality coding patterns seen at training time strongly influence the likelihood of generating low-quality code at inference time}.

All code, datasets and analysis scripts used in our study are publicly available for replication purposes \cite{replication}.
\section{Study Design} \label{sec:design}

The \emph{goal} of our study is to assess the impact of low-quality coding instances seen at training time by DL-based code generators on the quality of the code they produce. We aim at answering the following research questions (RQs):

\textbf{RQ$_1$:} \emph{To what extent training data collected by mining open source projects is affected by code quality issues?} This preliminary RQ assesses the extent to which code mined from open source projects and available in ready-to-use datasets is affected by quality issues. Siddiq \etal \cite{Siddiq:scam2022} reported between 39\% and 97\% of instances to be \emph{smelly} in the three datasets they investigated. However, these datasets were substantially smaller (between 117k and 251k functions) than the dataset we will employ in our study (4.4M functions). Thus, our differentiated replication will help in corroborating their findings.\smallskip

\textbf{RQ$_2$:} \emph{To what extent quality issues affecting the training instances impact the code generated at inference time?} This is the core RQ of our study in which we aim at understanding whether the low-quality instances seen at training time influence the quality of the code at inference time. Such an input-output relationship can be conjectured if (i) the model trained on a dataset featuring code quality issues actually generates low-quality code at inference time; and (ii) the same model trained on a cleaned version of the dataset (\ie one from which we remove the low-quality instances) has a statistically significant lower chance of generating low-quality code.\smallskip

In the following we detail the DL model employed in our study (\secref{sub:model}), and the process used to build the training, evaluation, and test sets (\secref{sub:datasets}) and to detect quality issues in them (\secref{sub:detecting}). Finally, \secref{sub:procedure} details the data collection and analysis procedure.

\subsection{Deep Learning Model}
\label{sub:model}
DL models used for code generation typically employ a decoder-only Transformer architecture~\cite{vaswani2017attention} to generate code based on a prompt. DeepSeek-Coder \cite{guo2024deepseek} is a prominent example of these models and reported state-of-the-art performance on code-related benchmarks among open code models \cite{guo2024deepseek}. As described by the authors, DeepSeek-Coder has been ``\emph{trained on a meticulously curated project-level code corpus, utilizing a fill-in-the-blank pre-training objective to enhance code infilling capabilities}'' \cite{guo2024deepseek}. The pre-training corpus of DeepSeek-Coder features 2 trillion tokens (87\% code, 10\% English text extracted from GitHub's Markdown and StackExchange, and 3\% high-quality Chinese articles). The code comes from projects written in 87 programming languages. 


In this work, we adopt the DeepSeek-Coder-Instruct-1.3B pre-trained model. The model has 24 hidden layers and uses a Switchable Gated Linear Unit (SwiGLU) hidden activation function to dynamically adjust activation pathways. The model leverages a multi-head attention mechanism with 16 attention heads. Further architectural details are available in \cite{guo2024deepseek}. While, as we will show, this model is able to support code generation (\ie taking as input a code description and implementing it), we will further fine-tune it for such a task by providing pairs of $\langle$description, code$\rangle$ (from now on, $\langle$\emph{d}, \emph{c}$\rangle$), where ``code'' is a Python function and ``description'' its English description. This allows to study how the presence (or lack) of low-quality instances in the fine-tuning training set influences its likelihood of generating low-quality code at inference time. We will also compare the fine-tuned models (\ie the one fine-tuned on the whole dataset and the one fine-tuned on \emph{clean} instances only) with the pre-trained-only model \cite{guo2024deepseek}, both in terms of performance (\ie ability of generating the target code) and code quality at inference time.

\subsection{Training, Evaluation, and Test Sets}
\label{sub:datasets}

To create our training, evaluation, and test sets we started from the The Stack \cite{thestack}, a publicly available dataset featuring 6TB of code written in 30 programming languages. We downloaded from it $\sim$24M files written in Python. The choice of focusing on Python derives from (i) its popularity, with it being the second most-popular language on GitHub \cite{pythonPopularity}; and (ii) the availability of a wide range of low-quality issues detection rules in a highly scalable static analyzer which can work on raw code --- details in \secref{sub:detecting}. 

We use the lizard library \cite{lizard} to extract from each Python file all functions declared in it. We remove from such a set all functions not having a \texttt{docstring}, since we need a functions' description to generate the above-mentioned $\langle$\emph{d}, \emph{c}$\rangle$ pairs. This resulted in $\sim$73M functions with their associated documentation collected. 

We then perform a cleaning process on the \texttt{docstring} aimed at removing noisy information that can confuse the model rather than guiding it in the code generation. For example, \texttt{docstring} may contain links to external resources, examples on how to use the code, and many other elements making them diverging from the kind of ``code description'' expected for this task. 

While most of the works in the literature adopt the first sentence of the \texttt{docstring} as a representative of a (short) code description \cite{Hu:icpc2018,Li:fse2020,Mastropaolo:icse2023}, we decided for a more thorough \texttt{docstring} cleaning process aimed at extracting all sentences actually composing the function description. Indeed, we found that most of collected \texttt{docstrings} ($\sim$70\%) feature multiple sentences describing the function's behavior. Thus, the second author manually inspected a random sample of \texttt{docstrings} from the dataset, defining heuristics to remove from them noisy information while keeping all sentences describing the function's behavior. The process was iterative and repeated several times by running the implemented heuristics on a sample of the dataset, inspecting the obtained output, and refining the cleaning pipeline until the result was satisfactory. 
To validate the quality of the collected data, we manually inspected a sample of $\langle$\emph{d}, \emph{c}$\rangle$ pairs to account for the presence of any incorrect or outdated docstrings, as detailed in~\secref{sec:threats}. 

\begin{figure}
    \centering
    \includegraphics[width=0.8\linewidth]{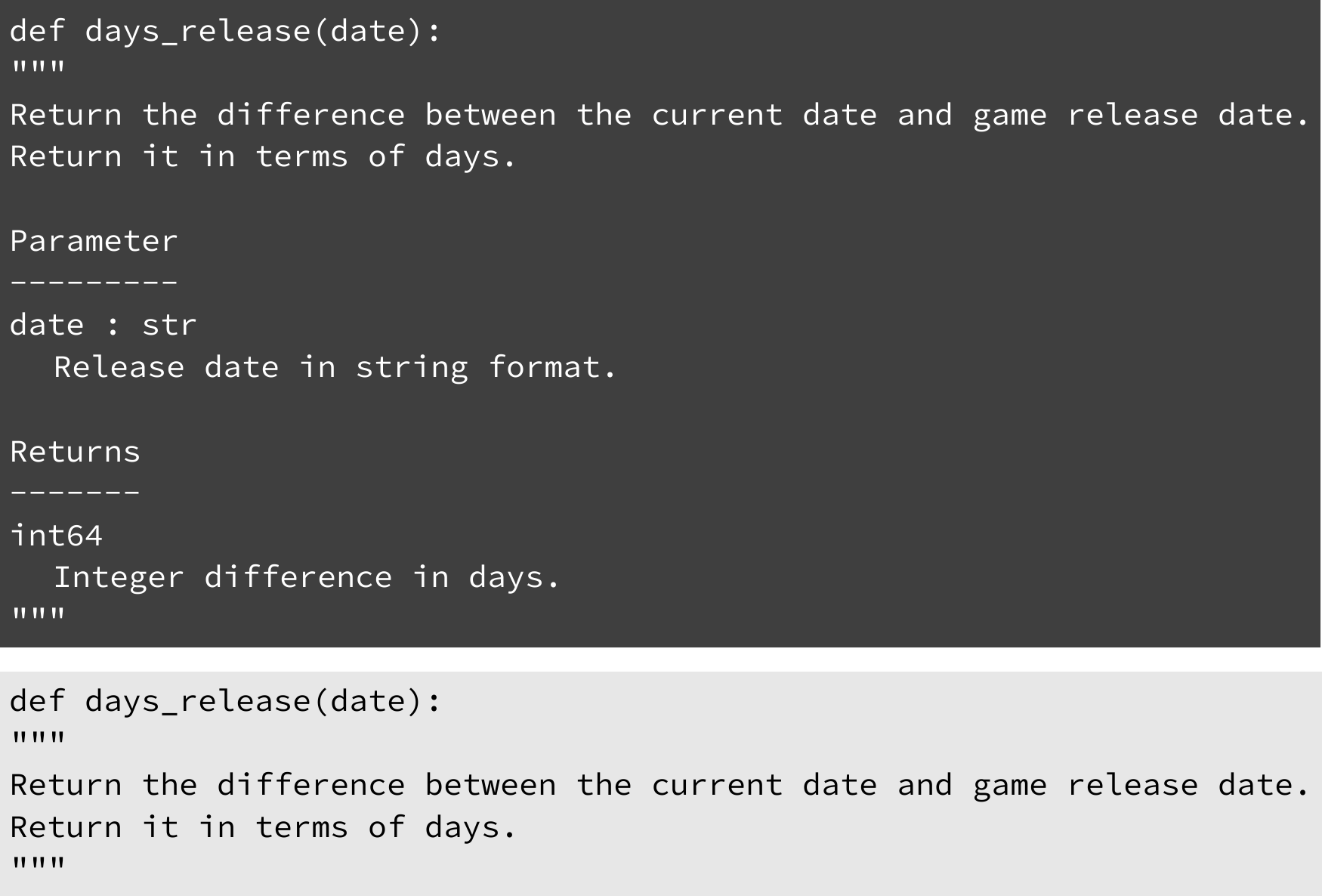}
    \vspace{-0.2cm}
    \caption{Docstring cleaning example.}
    \label{fig:cleaning}
    \vspace{-0.4cm}
\end{figure}

The obtained cleaning script is available in our replication package \cite{replication} and features several steps. Here we describe the main ones. We start by removing sentences documenting specific parts of the code rather than providing an overall description of the function (\ie descriptions of parameters and return values). These sentences are identified by matching specific patterns in the \texttt{docstring} (\eg ``\emph{Parameters:}'' or ``\emph{Returns:}''). 
This choice simulates a typical code generation scenario where developers provide a brief function description and signature, aligning with popular benchmarks \cite{chen2021evaluating}.
Similarly, code included in the \texttt{docstring} to showcase how to use the function is parsed and removed. Then, empty lines are removed and functions having a \texttt{docstring} including non-ASCII characters are excluded in an attempt to filter out $\langle$\emph{d}, \emph{c}$\rangle$ pairs having a non-English description. The ``description'' in the remaining pairs is then cleaned from special tags used to format the \texttt{docstring} (\eg `$<$summary$>$', `$<$note$>$'). Subsequently, we discard all descriptions including a link. Indeed, we found out that (i) links are generally used to point to additional information needed to understand the function and such information would not been available to the model used for code generation; and (ii) only 2\% of the remaining $\langle$\emph{d}, \emph{c}$\rangle$ pairs had a description including a link. Thus, discarding these pairs was a safer choice to ensure the quality of the dataset. Finally, $\langle$\emph{d}, \emph{c}$\rangle$ pairs featuring a description having less than 10 words have been excluded to remove too short descriptions unlikely to provide the model with a meaningful specification of the function to implement. After these steps, $\sim$8M $\langle$\emph{d}, \emph{c}$\rangle$ pairs are left. \figref{fig:cleaning} shows an example of \texttt{docstring} before (top) and after (bottom) the application of our cleaning pipeline. 

Besides cleaning the descriptions, we also cleaned the code. First, we removed inline comments, since this simplifies (i) the fine-tuning, guiding the model to only generate code; and (ii) the assessment of the correctness of the generated functions via quantitative metrics such as exact matches, \ie cases in which the function implemented by the model starting from a description is identical to the target (expected) one. Then, we format all functions using the \texttt{black} library \cite{black} to ``teach'' the model a consistent formatting style. Finally, we exclude pass functions, used in Python as placeholder for future code, and tests, identified as functions featuring the word ``\texttt{test}'' in their name. 
Excluding tests allows to focus the model on the generation of production code ($\sim$7M $\langle$\emph{d}, \emph{c}$\rangle$ pairs). 

The last step in the preparation of the datasets consists in excluding $\langle$\emph{d}, \emph{c}$\rangle$ pairs featuring a description and/or a function that is too long. Such a step is only needed to reduce the computational cost of training the model, as also done in previous work on code generation \cite{Cotroneo:icpc2024, yang2024intercode, ugare2024improving}.
While for descriptions we set the maximum length to 50 tokens (words and punctuation), applying the same token-level threshold on functions does not yield consistent results due to the differences in how the DeepSeek-Coder tokenizer processes code. To make a concrete example, we had two functions both composed by 150 Python tokens but resulting in 509 and 1,503 tokens, respectively, when processed by the DeepSeek-Coder's tokenizer. This difference is mostly due to the length of the identifiers used in the code. For example, two identifiers \texttt{a} and \texttt{THIS\-\_IS\-\_A\-\_CONSTANT} both represent a single Python token, but result in a different number of tokens when processed via DeepSeek-Coder. To account for this variability, we set a constraint on either the number of Python tokens or the number of characters, 450 Python tokens or 800 characters, respectively, with the aim of excluding functions likely to result in an excessive number of tokens.


The left 5,516,412 $\langle$\emph{d}, \emph{c}$\rangle$ pairs have finally been converted into $\langle$\emph{d+s}, \emph{c}$\rangle$ pairs, with ``s'' representing the signature of the function (``c'') to generate. Indeed, the model's input for the code generation task we experiment with will be the function description and signature, with the expected output being the complete function implementation. These pairs have been split into training (80\%), evaluation (10\%), and test (10\%) set as depicted in the ``Full dataset'' column of \tabref{tab:datasets}.

\begin{table}
    \centering
    \caption{Full and cleaned datasets used in our study\vspace{-0.2cm}}
    \begin{tabular}{lrr}
    	 \toprule
         & \textbf{Full Dataset} & \textbf{Cleaned Dataset} \\
         \midrule
         \textbf{Training} &  4,413,130 & 4,156,154 \\
         \hspace{0.5cm}\textit{low-quality} & 219,723 & - \\
         \hspace{0.5cm}\textit{syntactically incorrect} & 37,253 & - \\
         \hspace{0.5cm}\textit{clean} & 4,156,154  & 4,156,154 \\
         \midrule
         \textbf{Evaluation} & 551,641 & 551,641 \\
         \midrule
         \textbf{Test} & 551,641 & 551,641 \\
         \bottomrule
    \end{tabular}
    \label{tab:datasets}
    \vspace{-0.25cm}
\end{table}

\subsection{Detecting Quality Issues}
\label{sub:detecting}

To detect quality issues in the training set and, later on, in the code generated by the models at inference time, we employ Semgrep OSS~\cite{semgrep}. Semgrep is an industry-level, lightweight, static analysis tool for finding security vulnerabilities, bugs, and code quality issues in more than 30 programming languages. Python is the language for which the widest set of detection rules is available in Semgrep, leading to the most comprehensive set of quality issues supported. 

Semgrep is based on a pattern-matching detection approach which does not require code compilation, making it especially suited for scanning large-scale datasets with code coming from thousands of projects, as in our case. 
Semgrep offers a registry of rulesets collecting detection rules covering the same macro-category of issues (\eg a ruleset for detecting vulnerabilities belonging to the OWASP Top Ten~\cite{owasp}, a ruleset for detecting unencrypted communications, \etc). If a rule is present in multiple rulesets, it is run only once. In our investigation, we include all the rules available for Python in June 2024, excluding only rules for detecting AI-generated code. The final configuration for our analysis features a total of 686 rules (complete list available in our replication package \cite{replication}), spanning over six different categories of code quality issues: security (582), correctness (37), best-practice (24), compatibility (20), maintainability (16), and performance (7). For each identified quality issue, Semgrep returns: (i) the issue id (\ie the name of the rule that matched); (ii) the severity of the issue (\ie info, warning, or error); (iii) the category of the issue (\ie security, correctness, \etc); (iv) the line(s) of code affected by the issue; (v) an explanation of the detected problem; and (vi) for the security category, the mapping of the detected issue to the MITRE's Common Weakness Enumeration (CWE) \cite{CWEs}. To validate Semgrep's accuracy in correctly identifying quality issues, we conduct a manual inspection of a statistically significant random sample of issues detected in the code, as detailed in~\secref{sec:threats}.


\subsection{Experimental Procedure \& Data Analysis}
\label{sub:procedure}

To address \textbf{RQ$_1$}, we run Semgrep on the 4,413,130 Python functions in our training set (see \tabref{tab:datasets}), storing the list of identified quality issues. This is all we need for RQ$_1$ (\ie quantifying and characterizing the functions affected by quality issues). Running Semgrep also provides us with the set of functions on which syntax errors have been found. 

\eject

\tabref{tab:datasets} anticipates the number of \emph{low-quality}, \emph{syntactically incorrect}, and \emph{clean} functions in our training set, since those are needed to introduce later on the data collection for RQ$_2$. 

We answer RQ$_1$ by depicting a Sankey diagram \cite{Otto:2022} showing the percentage of quality issues identified in our training set. The diagram details the types of quality issues affecting the training instances, with a breakdown into security, correctness, best-practice, maintainability, compatibility, and performance. Also, for each of these categories, we highlight the top-5 most popular quality issues (\eg vulnerability types in the security categories), providing in the replication package the full list of identified quality issues. 

To address \textbf{RQ$_2$}, we use the full training dataset of $\sim$4.4M $\langle$\emph{d+s}, \emph{c}$\rangle$ pairs to fine-tune the DeepSeek-Coder model for the task of code generation. We refer to the model trained on the full dataset as DSC$_f$ (DeepSeek-Coder ``full''). As shown in \tabref{tab:datasets}, the full dataset features low-quality and syntactically incorrect functions (\ie the \emph{c} code to generate) and this allows to assess the extent to which these instances influence the chances that DSC$_f$ will generate low-quality code at inference time (\ie when asked to generate the 551,641 functions in our test set). This requires five steps. First, we generate the 551k functions using DSC$_f$. 

Second, we run Semgrep on the functions outputted by DSC$_f$, identifying those affected by quality issues as well as the ones syntactically incorrect. Third, we create the ``cleaned'' version of our training set (see \tabref{tab:datasets}) by removing from the full dataset all low-quality and syntactically incorrect instances, obtaining a training set featuring 4,156,154 $\langle$\emph{d+s}, \emph{c}$\rangle$ pairs. Fourth, we fine-tune another DeepSeek-Coder model (DSC$_c$) on such a cleaned dataset. Finally, we use DSC$_c$ to generate the same 551k functions previously synthesized by DSC$_f$. This allows to compare the percentage of low-quality and syntanctically incorrect functions generated by a model trained on the full dataset (DSC$_f$) and one trained on a cleaned version of it (DSC$_c$). To do that, we report the percentage of low-quality functions generated by the two models, also detailing the type of quality issues affecting these functions. 

While not really needed to answer RQ$_2$, we also report the percentage of low-quality functions generated by the pre-trained only model (from now on, ``PTO''), since it is interesting to see the percentage of low-quality functions generated by a publicly available state-of-the-art code generator. Clearly, not having access to the pre-training dataset used by the authors of DeepSeek-Coder \cite{guo2024deepseek} we cannot investigate the extent to which possible low-quality code it will generate may have been influenced by the coding patterns learnt at training time. The investigation of such an ``input-output relationship'' (RQ$_2$) is left to the above-described comparison between the predictions generated by DSC$_f$ and DSC$_c$. 

We also statistically compare the percentage of low-quality functions generated by PTO, DSC$_f$, and DSC$_c$. We assume a significance level of 95\%. The compared distributions both feature 551,641 elements (one for each function to be generated in the test set) and a boolean value indicating whether the generated function is affected or not by a quality issue. As explained below, we use non-parametric tests since the compared distributions deviate from normality according to the Anderson-Darling test \cite{adtest} (\emph{p}-values$<$0.001). To (pairwise) compare the low-quality functions generated by the three models, we use the McNemar's test \cite{mcnemar}, which is a proportion test suitable to pairwise compare dichotomous results of two  treatments. To compute the test results (\eg for DSC$_f$ \emph{vs} DSC$_c$), we create a confusion matrix counting the number of cases in which (i)  both DSC$_f$ and DSC$_c$ generate a low-quality function, (ii) only DSC$_f$ generates a low-quality function, (iii) only DSC$_c$ generates a low-quality function, and (iv) neither DSC$_f$ nor DSC$_c$ generates a low-quality function. 

We complement the McNemar's test with the Odds Ratio (OR) effect size. We account for multiple test instances (\ie comparing DSC$_f$ with both PTO and DSC$_c$), by adjusting \emph{p}-values using the Benjamini-Hochberg procedure \cite{yoav:jstor1995}. The same analyses described for the generated low-quality functions across the three models are also presented for the percentage of syntactically wrong functions they generate.

While the data extracted allows to answer our research questions, it is also important to assess the performance the DeepSeek-Coder models (PTO, DSC$_f$, and DSC$_c$) for the task of interest (code generation). 

Indeed, showing that the functions implemented by a model feature quality issues would be of little interest for a model that cannot deal with the task (\eg a model outputting random code). This analysis also helps to ensure that the fine-tuning procedure we performed did not decrease the performance of the PTO model for the code generation task. To assess the models' performance, we (i) compute metrics acting as proxies of code correctness on the predictions generated by the experimented models on our large-scale test set; and (ii) run the models on an additional (smaller) test set featuring test cases, thus allowing a more reliable assessment of code correctness for the generated functions. 

Concerning the first point, we compute the number of Exact Match (EM) predictions and the CrystalBLEU score \cite{Eghbali:ase2023} between the predictions made by the three models  and the target function. An EM indicates that the output of the model is identical to the expected target. While an EM guarantees the correctness of the generated function, it represents a very strict definition of correctness, considering that the same function may be correctly implemented in several different ways. For this reason, we complement our analysis with the CrystalBLEU score \cite{Eghbali:ase2023}, which is a variant of the BLEU score~\cite{papineni:acl2002} tailored to better assess similarity among code snippets. The BLEU score is based on the overlap of n-grams between the prediction and the target. The basic idea behind CrystalBLEU is to ignore from such a computation the most frequent n-grams in a language. Indeed, in the case of programming languages, these frequent n-grams are mostly due to the syntactic constructs and coding conventions of the language, thus inflating and flattening the computed similarity values. As shown by Enghbali and Pradel \cite{Eghbali:ase2023}, CrystalBLEU can discriminate similar from dissimilar code up to 4.5 times more effectively as compared to the original BLEU score. We statistically compare the EM predictions generated by the models using again McNemar's test \cite{mcnemar} with the OR effect size. As for the CrystalBLEU, we use the paired Wilcoxon signed-rank test \cite{wilcoxon} and the Cliff's delta \cite{Cliff:2005} effect size. Also in this case we account for multiple tests by adjusting \emph{p}-values using the Benjamini-Hochberg procedure \cite{yoav:jstor1995}.

As per the second point, we run the three models on the HumanEval benchmark~\cite{chen2021evaluating}, a dataset of 164 hand-written Python problems designed to evaluate the performance of AI code generators. Each problem includes a function signature, docstring, body, and a set of unit tests to verify the functional correctness of the generated code. 

Also in this case, besides reporting the percentage of functions correctly generated by each model (pass@1), we statistically compare them using the McNemar's test \cite{mcnemar} (with adjusted \emph{p}-values) and the OR.

\subsubsection{Training procedure} We detail here the technicalities behind our training (fine-tuning) procedure followed to create DSC$_f$ and DSC$_c$. Models have been fine-tuned using 3 NVIDIA GPU GeForce RTX 3090 with 24 GB of RAM each. We use the default DeepSeek-Coder hyperparameters recommended by the authors \cite{guo2024deepseek}, using AdamW~\cite{loshchilov2017decoupled} as the optimizer with $\beta_1$ and $\beta_2$ values of 0.9 and 0.95, a learning rate of $2e^{-5}$ with \textit{cosine} scheduler type, and 10 warm-up steps. 

During training, we set to 1,024 the maximum number of tokens that can be processed between input (description and signature) and output (complete function). We trained both models for one epoch, for a total of 30,646 (DSC$_f$) and 28,861 (DSC$_c$) training steps using batch size of 12 and 4 accumulation steps. The lower number of training steps for DSC$_c$ is due to the smaller training set used (see \tabref{tab:datasets}).

During training, we saved a checkpoint and evaluated the model's performance on the validation set at every 20\% step on the training set (\ie after 20\% of the training dataset were seen, then after 40\%, \etc), for a total of 5 evaluation steps. We then selected the best checkpoint as the one having the lowest evaluation loss. Following this procedure, a single training took $\sim$48 hours. The best checkpoint was the second to last for DSC$_f$ and the last one for DSC$_c$. In both cases we observed a flattening of the loss curve starting from the third checkpoint (graph available in our replication package \cite{replication}).

\section{Results Discussion} \label{sec:results}


\begin{figure*}
    \centering
    \includegraphics[width=0.905\linewidth]{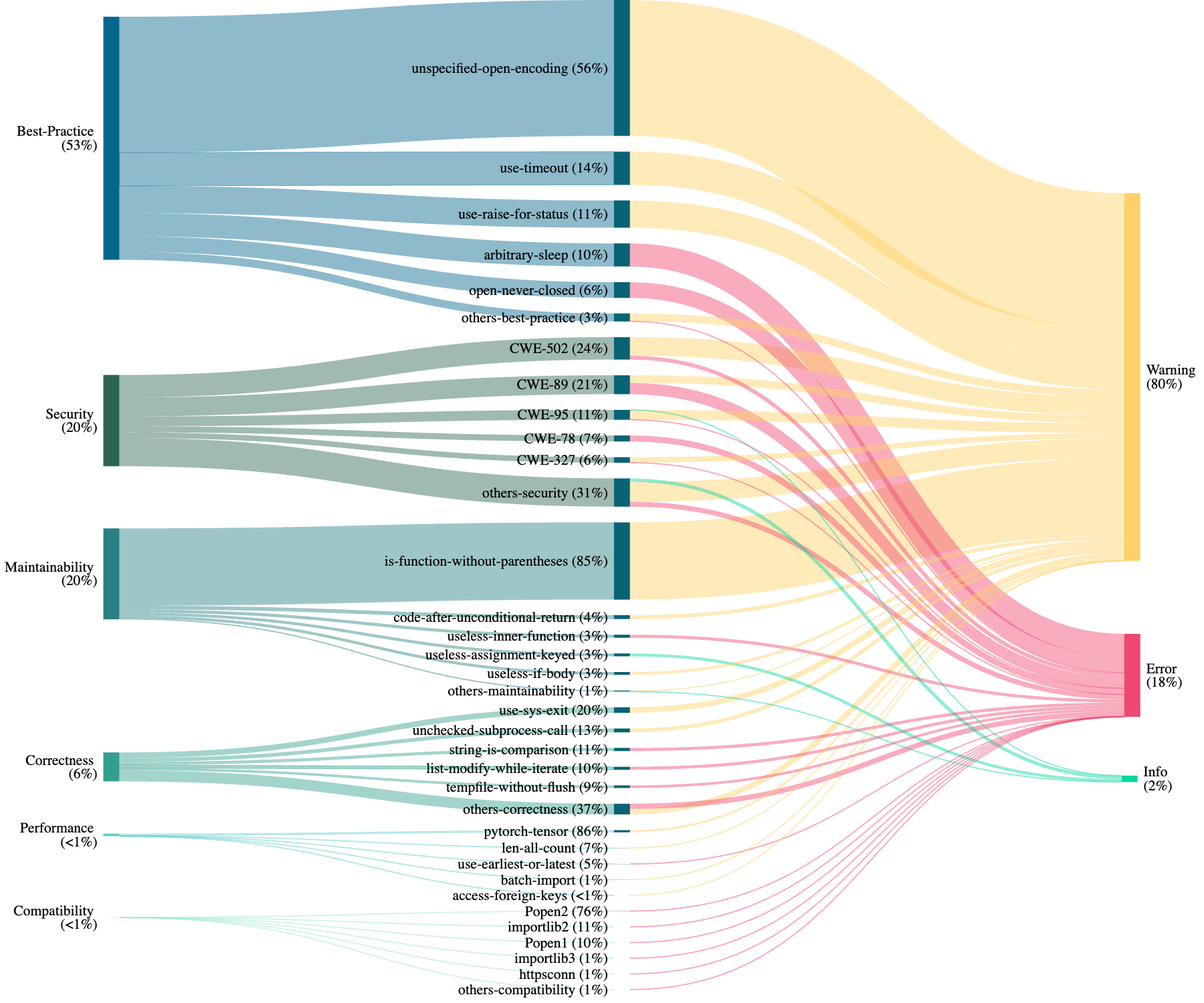}
    \caption{\vspace{-0.8cm}Types of quality issues found in the 4.98\% (219,723) low-quality functions in the fine-tuning dataset}
    \label{fig:sankey}
    \vspace{-0.4cm}
\end{figure*}

\subsection{RQ$_1$: To what extent training data collected by mining open source projects is affected by code quality issues?}

Semgrep found 219,723 low-quality functions in the training set we used for DSC$_f$ (4.98\% of the $\sim$4.4M functions). On top of that, 37,253 functions (0.84\%) have been reported as syntactically incorrect. Also, some of the low-quality functions are affected by multiple quality issues: Semgrep reported a total of 310,051 issues, resulting in an ``issue density'' of 1.4 issues per low-quality function. The Sankey diagram in \figref{fig:sankey} depicts the distribution of quality issues in the affected functions, detailing for each ``root'' category (\ie best-practice, security, maintainability, correctness, performance and compatibility) the top-5 quality issues in terms of frequency. 

For example, 20\% of the 219,723 low-quality functions are affected by security issues and, out of these, 24\% are related to the CWE-502 vulnerability (\ie deserialization of untrusted data). Finally, the right part of \figref{fig:sankey} classifies the quality issues into three possible groups based on their severity: (i) \emph{information}, indicating low-risk problems that do not affect code execution (\eg maintainability issues like dead-code); (ii) \emph{warnings}, pointing to issues possibly resulting in misbehaviors of the program (\eg an HTTP request without timeout causing the program to hang) or to medium-impact software  security vulnerabilities (\eg sensitive information stored in log files); and (iii) \emph{errors}, being bugs very likely to result in failures and high-risk security flaws (\eg exposure to path traversal attacks). As it can be seen, the majority of quality issues identified by Semgrep are warnings (80\%), followed by errors (18\%) and information ($<$2\%). 

Violations of Python best-practices constitute more than half of quality issues in the training set ($\sim$53\%), with warnings about not specifying the encoding when opening a file (\emph{unspecified\--open\--encoding} in \figref{fig:sankey}) representing the most frequent concern in this category. 

Since not all systems use \texttt{UTF-8} encoding by default, this inattention may lead to portability issues, data corruption or decoding runtime errors~\cite{python-unspecified-open-encoding}. Other popular issues concern the incorrect handling of HTTP requests, \ie not setting a timeout (\emph{use-timeout}) and not checking the status code in case of a failed request (\emph{use\--raise\--for\--status}). Best-practices issues can also have quite serious consequences, as it is the case for the fourth (\emph{arbitrary-sleep}) and fifth (\emph{open-never-closed}) most popular issues in this category. 
The former indicates the usage of fixed sleep intervals, mostly as a workaround for synchronization or timing issues and potentially causing program inefficiency, unpredictability or synchronization issues (\eg deadlocks). The latter relates to resources (\eg files) opened but not closed, with possible consequences in terms of resource leaks, performance degradation, and memory exhaustion.
The second most occurring category of issues in the training data relates to security vulnerabilities, which affect 20\% of low-quality functions.  The most spread vulnerability (see \figref{fig:sankey}) is \emph{CWE-502: Deserialization of Untrusted Data}, allowing attackers to exploit weak deserialization mechanisms in the code to perform unauthorized actions.  

The other popular security weaknesses in the top-5 concern injection attacks: SQL Injection (\emph{CWE-89}), Eval Injection (\emph{CWE-95}), and OS Command Injection (\emph{CWE-78}). These defects expose the software to data breaches, unauthorized access, and system compromise. 
Finally, \emph{CWE-327} is the last vulnerability in the top-5 and relates to the use of a broken or risky cryptographic algorithm (\eg MD5, SHA1). 
Interesting is also to note that the severity assigned by Semgrep for issues belonging to the same vulnerability type (\eg \emph{CWE-89}) varies based on the specific context in which the flaw occurs. This is due to estimates made by Semgrep about the likelihood of exploiting the vulnerability (\eg based on the reachability and location of the vulnerable statement).
Coming to maintainability issues, affecting $\sim$20\% of the low-quality functions, most of them concerns functions referenced without parentheses (\eg \texttt{return self.\-is\_ready}). Since functions in Python are objects, this syntax is correct but it should only be used in specific circumstances (\eg to pass the function as reference). Other maintainability issues in the top-5 concern dead/useless code.

Correctness issues are relatively rare, affecting $\sim$6\% of the low-quality functions, and they are mostly due to incorrect program termination (\emph{use-sys-exit}), calling sub-processes without checking the return status (\emph{unchecked\--subprocess\--call}), incorrectly comparing strings using \texttt{is} instead of '==' (\emph{string\--is\--comparison}), modifying a list while iterating over it (\emph{list\--modify\--while\--iterate}), and not flushing or properly closing a temporary file before using it (\emph{tempfile\--without\--flush}). These flaws can lead to unpredictable program behavior, unexpected results, data corruption, or runtime errors. 
Finally, $<$2\% of issues concern performance and compatibility issues (\eg using outdated/deprecated libraries). The whole list of detected quality issues is available in the replication package~\cite{replication}. 

Before concluding, it is worth discussing the difference between our findings (\ie 4.98\% of considered Python functions are affected by quality issues) and what found by Siddiq \etal \cite{Siddiq:scam2022}, who reported between 39\% and 97\% of the Python functions in the three inspected datasets (\ie Code Clippy, APPS, and CodeXGLUE) to be affected by at least one quality issue. This discrepancy can be attributed to the static analyzer used in their work \ie PyLint~\cite{pylint}. Indeed, they found that the most frequent issues affecting the subject functions concern bad code indentation and code lines being too long. Both these issues have been inherently excluded from our study as we ensure that all functions in our dataset are correctly formatted using the Python black library \cite{black}. 

To summarize, our investigation highlights that \emph{a non-negligible portion of functions ($\sim$5\%) in publicly available open source code datasets is affected by quality issues}. Our RQ$_2$ will look into the possible impact that this low-quality training data has on the code produced at inference time.

\begin{table}[tb]
    \vspace{-0.25cm}
    \centering
    \caption{Models' performance in code generation\vspace{-0.3cm}}
    \label{tab:metrics}
    \begin{tabular}{l r r r r r}
        \toprule
        \multirow{2}{*}{} & \multirow{2}{*}{\textbf{pass@1}} & \multirow{2}{*}{\textbf{EM}} & \multicolumn{3}{c}{\textbf{CrystalBLEU}} \\ \cline{4-6}
        & & & \textbf{mean} & \textbf{median} & \textbf{sd} \\ \midrule
        PTO        & $0.39$ & $0.03$ & $0.19$ & $0.11$ & $0.23$ \\ 
        DSC$_f$         & $0.51$ & $0.11$ & $0.29$ & $0.17$ & $0.31$ \\ 
        DSC$_c$ & $0.50$ & $0.11$ & $0.29$ & $0.17$ & $0.31$ \\ \bottomrule
    \end{tabular}
    \vspace{-0.4cm}
\end{table}

\subsection{RQ$_2$: To what extent quality issues affecting the training instances impact the code generated at inference time?}

Before discussing the prevalence of quality issues in the code generated by the experimented models, it is important to verify their performance to make sure they are representative of state-of-the-art DL-based code generation. \tabref{tab:metrics} reports (i) the pass@1 achieved on the HumanEval benchmark, namely the percentage of 164 Python functions part of the benchmark that the models correctly implemented; (ii) the EM predictions on the large-scale test set we built featuring $\sim$551k functions; and (iii) descriptive statistics for the CrystalBLEU, also in this case computed for the large-scale test set. 
\tabref{tab:statPerformance} reports the related statistical tests.

\vspace{-0.2cm}

\begin{table}[h!]
	\centering
	\caption{Statistical tests: Models' performance in code generation\vspace{-0.2cm}}
        \label{tab:statPerformance}
	\begin{tabular}{lrr}
		\toprule
		\multicolumn{3}{c}{\textbf{HumanEval: pass@1}}\\\midrule
		\textbf{Test} & \textbf{adj. \emph{p}-value} & \textbf{OR} \\
		\midrule
		DSC$_c$ \emph{vs} DSC$_f$ & 0.688 & -\\
		PTO \emph{vs} DSC$_c$ & 0.029 & 2.11\\
		PTO \emph{vs} DSC$_f$ & 0.029 & 2.00\\\midrule\\

		\toprule
		\multicolumn{3}{c}{\textbf{Large scale test set: EM}}\\\midrule
		\textbf{Test} & \textbf{adj. \emph{p}-value} & \textbf{OR} \\\midrule
		DSC$_c$ \emph{vs} DSC$_f$ & $<$0.001 & 1.29\\
		PTO \emph{vs} DSC$_c$ & $<$0.001 & 64.18\\
		PTO \emph{vs} DSC$_f$ & $<$0.001 & 61.10\\\midrule\\
		
		\toprule
		\multicolumn{3}{c}{{\bf Large scale test set: CrystalBLEU}}\\\midrule
		\textbf{Test} & \textbf{adj. \emph{p}-value} & \textbf{$|$\emph{d}$|$} \\
		\midrule
		DSC$_c$ \emph{vs} DSC$_f$ & $<$0.001 & 0.00\\
		PTO \emph{vs} DSC$_c$ & $<$0.001 & 0.22\\
		PTO \emph{vs} DSC$_f$ & $<$0.001 & 0.22\\
		\bottomrule
	\end{tabular}
	\vspace{-0.08cm}
\end{table}

First thing worth discussing is the effectiveness of the fine-tuning we performed. Indeed, both the model fine-tuned on the full (DSC$_f$) and on the cleaned (DSC$_c$) dataset achieved significantly better performance than the pre-trained only (PTO) model. This finding is confirmed across all metrics. For example, in terms of pass@1 on the HumanEval benchmark, the fine-tuned models managed to correctly implement 51\% (DSC$_f$) and 50\% (DSC$_c$) of the functions, against the 39\% of PTO. The statistical tests confirm (i) no significant difference in pass@1 between the fine-tuned models (adj. $p$-value = 0.688), and (ii) significantly higher pass@1 for the fine-tuned models with respect to PTO (adj. $p$-value = 0.029) with a ORs $\geq$ 2. Similar results can be observed in terms of EM predictions and CrystalBLEU values. Worth mentioning is that the EM predictions are in general quite low even for the best-performing models (11\%). As explained in \secref{sub:procedure}, this is due to the fact that a prediction is an EM only if it is identical token by token to the target. A single different token between the prediction and the target (\eg an identifier named \texttt{num} rather than \texttt{number}) does not result in an EM. 

Thus, it is still quite impressive that the fine-tuned models managed to generate $\sim$ 0.11$\times$551k=61k functions identical to the target. Also, note that the statistically significant difference observed between DSC$_f$ and DSC$_c$ in terms of EM predictions and CrystalBLEU (see \tabref{tab:statPerformance}) is mostly due to the vast amount of data (551k elements) in the two distributions, as also shown by the quite low OR (1.29) when comparing their EMs, and the 0.00 (negligible) Cliff's delta ($d$) when comparing their CrystalBLEU scores. Differently, ORs are quite high ($>$ 60) when comparing the EMs achieved by the fine-tuned models with those of PTO, and the Cliff's $d$ is 0.22 (small) when focusing on the CystalBLEU scores. 

In short, according to all our metrics, \emph{the applied fine-tuning significantly boosted the performance of the PTO model for the task of code generation}. Instead, \emph{no major differences are observed between the model fine-tuned on the full dataset, and the one fine-tuned on its cleaned version.}

Besides the observed improvement, it is also important to verify if the performance achieved by the fine-tuned models is representative of state-of-the-art code generators. By looking at previous work comparing several DL models on the same HumanEval benchmark \cite{yu2024codereval,liu2024your}, we found that the fine-tuned DeepSeek-Coder-1.3B models achieved better/similar results as compared to other, larger, open code models. For example, DSC$_f$ achieved 51\% of pass@1, against the 52\% achieved by CodeLlama 34B \cite{roziere2023code} and the 34\% of StarCoder 15B \cite{li2023starcoder}. When looking at LLMs having trillions of parameters, results reported in the literature vary depending on the adopted models, but are still not very far from what DeepSeek-Coder achieved. For example, for ChatGPT the reported pass@1 spans between 39\% (GPT-3.5-Turbo \cite{liu2024your}) and 69\% (GPT-4 \cite{liu2024your}). Thus, the \emph{fine-tuned models can be considered representative of DL models supporting code generation}.

\vspace{-0.15cm}

\begin{table}[h!]
    \centering
    \caption{Percentage of syntactically incorrect and low-quality functions generated by PTO, DSC$_f$, and DSC$_c$\vspace{-0.3cm}}
    \label{tab:lowQuality}
    \begin{tabular}{l r r r}
        \toprule
        & \textbf{PTO} & \textbf{DSC$_f$} & \textbf{DSC$_c$} \\ \midrule
        \textbf{Syntact. incorrect}             & $15.45\%$ &  $7.13\%$ &  $6.68\%$ \\ \midrule
        \textbf{Low-quality}                    &  $7.09\%$ &  $5.85\%$ &  $2.16\%$ \\ \midrule
        \textbf{Total \# of issues}                   &  $62,715$ & $73,251$ &  $22,466$ \\
        \noalign{\vskip 0.1cm}
        \hspace{0.5cm}\textit{maintainability}  & $44.38\%$ & $65.37\%$ & $61.27\%$ \\ 
        \hspace{0.5cm}\textit{best-practice}    & $45.57\%$ & $25.10\%$ & $20.58\%$ \\ 
        \hspace{0.5cm}\textit{security}         &  $8.21\%$ &  $6.87\%$ & $12.29\%$ \\ 
        \hspace{0.5cm}\textit{correctness}      &  $1.79\%$ &  $2.56\%$ &  $5.71\%$ \\ 
        \hspace{0.5cm}\textit{performance}      &  $0.05\%$ &  $0.05\%$ &  $0.08\%$ \\ 
        \hspace{0.5cm}\textit{compatibility}    &  $0.00\%$ &  $0.05\%$ &  $0.08\%$ \\ 
        \bottomrule
    \end{tabular}
    \vspace{-0.2cm}
\end{table}

Moving to the core of our research question, \tabref{tab:lowQuality} reports the percentage of syntactically incorrect and low-quality functions (\ie functions featuring at least one quality issue identified by Semgrep) generated by PTO, DSC$_f$, and DSC$_c$ in our test set (551k instances). \tabref{tab:lowQuality} also reports the total number of low-quality issues and their distribution by type in the generated functions. \tabref{tab:statQuality} complements this analysis with the results of the statistical tests comparing the proportions of low-quality and syntactically wrong functions in the predictions generated by the three models.

Concerning the functions generated by DSC$_f$ (\ie the model trained on the full dataset featuring 4.98\% of low-quality functions and 0.84\% of syntactically incorrect functions), 5.85\% of them ($\sim$33k) feature at least one quality issue, with additional 7.13\% ($\sim$39k) being syntactically incorrect. While there seems to be a ``proportion'' between the low-quality functions in the training set and those generated at inference time, the percentage of syntactically incorrect functions is substantially higher among the AI-generated functions. This is a quite expected result considering that the DL models are known to make some mistakes during the code generation.

As per the type of quality issues injected by DSC$_f$, the main observation that can be made is that, while the \emph{best-practice} type accounts for 53\% of quality issues in its training set, it represents ``only'' 25\% of the issues in the generated code, with the dominant issue type becoming \emph{maintainability} (65\% \emph{vs} 20\% of the training set). The tendency of the model to inject a higher proportion of \emph{maintainability} issues is mostly due to two types of issues which are frequent in the AI-generated functions but rarely present in the training set. The first concerns the tendency of DSC$_f$ to omit parentheses from a method call, an issue already discussed in RQ$_1$. The second relates to useless \texttt{if} statements written by the model (\eg the \texttt{if} and the \texttt{else} body feature the same instructions). These issue types are rarely found in human-written code. 

\vspace{-0.2cm}

\begin{table}[h!]
	\centering
	\caption{Statistical tests: Syntactically incorrect and low-quality functions generated by PTO,  DSC$_f$, and DSC$_c$\vspace{-0.3cm}}
        \label{tab:statQuality}
	\begin{tabular}{lrr}
		\toprule
		\multicolumn{3}{c}{\textbf{Syntactically incorrect}}\\\midrule
		\textbf{Test} & \textbf{adj. \emph{p}-value} & \textbf{OR} \\\midrule
		DSC$_c$ \emph{vs} DSC$_f$ & $<$0.001 & 1.14\\
		DSC$_f$ \emph{vs} PTO & $<$0.001 & 2.46\\
		DSC$_c$ \emph{vs} PTO & $<$0.001 & 2.63\\\midrule\\

		\toprule
		\multicolumn{3}{c}{{\bf Low-quality}}\\\midrule
		\textbf{Test} & \textbf{adj. \emph{p}-value} & \textbf{OR}\\
		\midrule
		DSC$_c$ \emph{vs} DSC$_f$ & $<$0.001 & 6.37\\
		DSC$_f$ \emph{vs} PTO & $<$0.001 & 1.32\\
		DSC$_c$ \emph{vs} PTO & $<$0.001 & 4.22\\
		\bottomrule
	\end{tabular}
	\vspace{-0.2cm}
\end{table}

More interesting is the comparison between the quality of functions generated by DSC$_f$ and those outputted by DSC$_c$, the model trained on the cleaned dataset. The percentage of low-quality functions generated by DSC$_c$ is less than half of those generated by DSC$_f$ (2.16\% \emph{vs} 5.85\%). This difference is statistically significant (adj. $p$-value $<$ 0.001) and accompanied by an OR=6.37, indicating six times higher odds of obtaining a low-quality function using the DSC$_f$ model rather than DSC$_c$. This is the main finding of our work: \emph{low-quality coding patterns seen at training time strongly influence the likelihood of generating low-quality code at inference time}. 

We also observed that the low-quality functions generated by the two models tend to exhibit a different \emph{density} of quality issues: Each low-quality function generated by DSC$_f$ is affected, on average, by 2.3 quality issues, while those generated by DSF$_c$ exhibit, on average, 1.9 quality issues. 

Consequently, DSC$_f$ generates $\sim$73k quality issues, compared to the $\sim$22k generated by DSC$_c$, representing more than a threefold decrease. Interesting is also to see that, despite the smaller training set used for DSC$_c$ (-257k instances as compared to DSC$_f$), this model tends to make less syntactical errors as compared to DSC$_f$ (6.68\% \emph{vs} 7.13\%, \ie -2.5k syntactically wrong functions). Remember that from the cleaned dataset we removed the 0.84\% of syntactically incorrect functions present in the full dataset. Such a cleaning seems to have played a role.

\begin{figure}
    \centering
    \label{fig:qualitative}
    \includegraphics[width=0.8\linewidth]{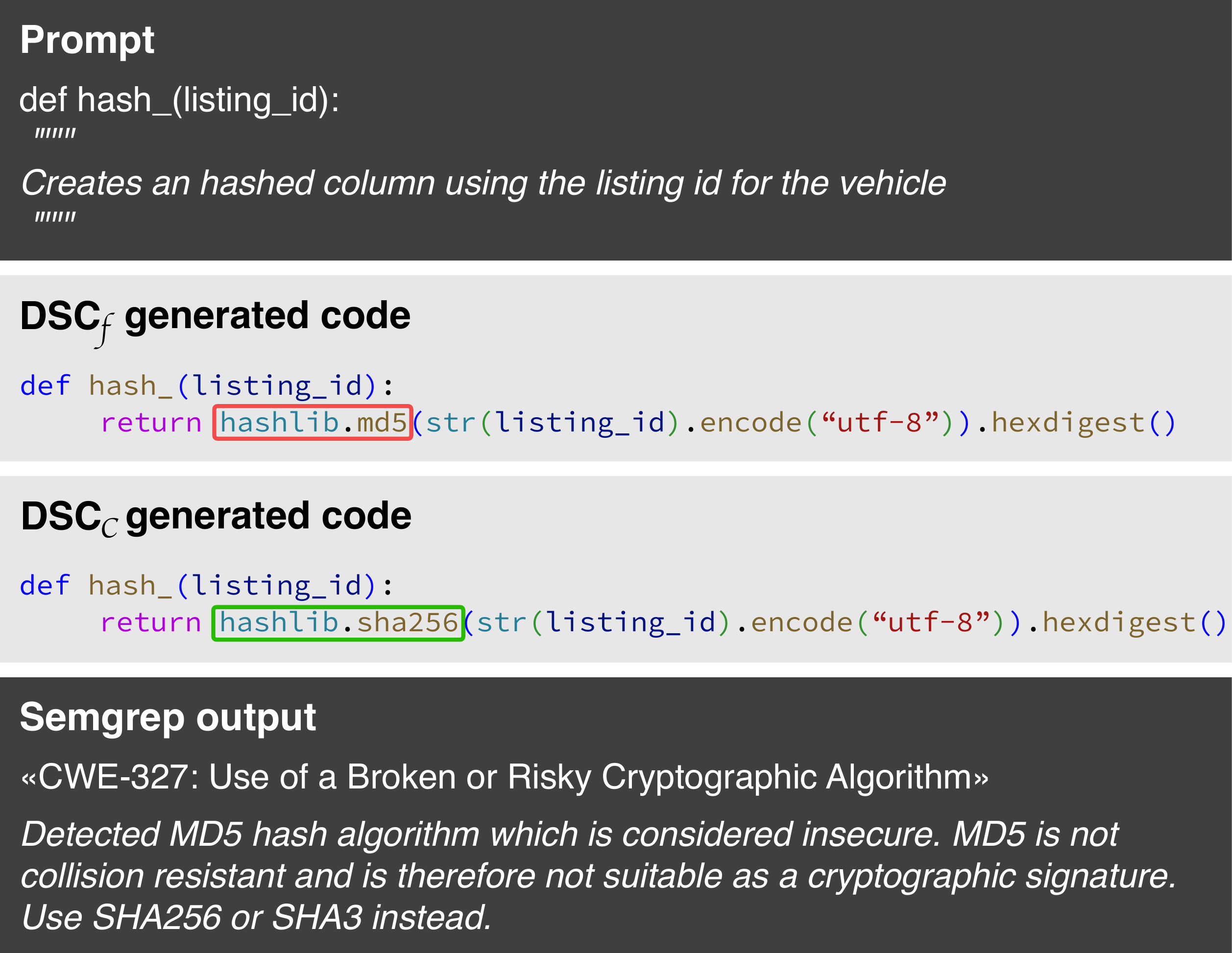}
    \caption{Functions generated by DSC$_f$ and DSC$_c$ for the same prompt}
    \vspace{-0.2cm}
\end{figure}

\figref{fig:qualitative} shows two simple functions implemented by DSC$_f$ and DSC$_c$ starting from the same prompt in our test set. The prompt requires the implementation of a function that \emph{creates an hashed column using the listing id for the vehicle}, where \texttt{listing\_id} is a function's parameter. Both implemented functions are functionally correct (\ie they meet the requirements specified in the prompt). However, for the one implemented by DSC$_f$, Semgrep reports the issue shown at the bottom of the figure and pointing to the low security of the \texttt{MD5} hash algorithm used in the function. The error message recommends the usage of \texttt{SHA256} or \texttt{SHA3} instead, with the former being the one used by the DSC$_c$ model.

Finally, it is worth commenting on the low-quality and syntactically incorrect functions generated by PTO, the pre-trained only DeepSeek-Coder model which is publicly available for download \cite{deepseekHF}. Such a model generates the highest percentage of low-quality (7.09\%) and syntactically incorrect (15.45\%) functions. Without having access to the pre-training dataset used by the original authors, it is difficult to make claims about the reasons for such results. Maybe they are simply due to the lack of fine-tuning for the specific task of code generation. However, even assuming that the low-quality functions are due to low-quality (pre-)training data, our findings suggest that a fine-tuning on ``clean'' code is recommended for the code generation task, since resulting in (i) better performance (\tabref{tab:metrics}) and (ii) higher-quality code (\tabref{tab:lowQuality}).
\section{Threats to Validity} \label{sec:threats}

\emph{Construct validity}. We used the \texttt{docstring} extracted from the Python functions as the prompt for code generation. 

While this is a standard practice in experiments related to code generation/summarization \cite{Hu:icpc2018,Li:fse2020,Mastropaolo:icse2023}, these descriptions may not be fully representative of prompts developers would write when asking DL models to generate code. To validate dataset quality, the first two authors independently rated 200 $\langle d, c \rangle$ pairs on \emph{content adequacy} (how well the comment summarizes the code) and \emph{understandability} (ease of comprehension) using a 3-point scale (1 = low, 3 = high). Disagreements occurred in 10.5\% of cases for content adequacy and 12.5\% for understandability, all being 1-point differences. Conflicts were resolved through discussion with a third author. The average scores were 2.84 for content adequacy and 2.81 for understandability, with only four summaries rated 1.

Another threat is related to the metrics used to assess the performance of the models on our large-scale test set. As said, EM predictions are a lower-bound for the actual correct predictions generated by the models, while CrystalBLEU is a similarity metric mostly indicating how close is the prediction to the expected code. To account for these limitations, we also assessed the code generation capabilities of the models on the HumanEval benchmark~\cite{chen2021evaluating}, showing that the fine-tuned models achieved a pass@1 score superior to that of bigger models (\eg CodeLLaMa, StarCoder, GPT-3.5-turbo) \cite{yu2024codereval, liu2024your}. This supports the validity of our training procedure.

Finally, we used Semgrep \cite{semgrep} to identify quality issues in both training code and model-generated functions. Semgrep does not require compilation and it is built around textual patterns matched in the code, posing questions about its accuracy. To account for that, the first two authors (6 and 8 years of programming in Python, respectively) manually inspected 400 randomly sampled quality issues, classifying each detected issue as a true or a false positive. Such a sample is statistically significant ensuring a 95\% confidence level $\pm$5\% \cite{Rosner2011}. 
They disagreed on 2.25\% of the instances. Conflicts have been solved by a third author. 
Ultimately, 97.75\% of reported issues were true positives, minimizing concerns about detection accuracy affecting our findings.

\emph{Internal validity}. We adopted the standard configuration of DeepSeek-Coder without performing hyperparameters tuning. However, the goal of our study was not to present the best-in-class code generator, but rather to study the impact of low-quality training data on the quality of the generated code. Still, we acknowledge that the specific configuration used may have played a role in our findings.


\emph{External validity}. While our study is large-scale when looking at the number of functions we generated with the subject models ($\sim$551k), our findings are limited to one model (DeepSeek-Coder), one language (Python), and one dataset  \cite{thestack}. At least we (i) adopted a model known in literature \cite{guo2024deepseek} for being well-representative of DL-based code generators (as also confirmed by our experiment on the HumanEval benchmark); and (ii) employed a large-scale dataset featuring millions of functions mined from GitHub projects and, thus, likely to be representative of code mined in the wild for the training of DL models. Still, our findings should only be considered valid in our specific study context.
\section{Related Work} \label{sec:related}


Progress in the field of AI-based code generation is fast-paced, with dozens of new models presented in the last couple of years (see \eg CodeLLama~\cite{roziere2023code}, AlphaCode~\cite{li2022competition}, DeepSeek-Coder~\cite{guo2024deepseek}, PaLM-Coder~\cite{chowdhery2023palm}, Starcoder~\cite{li2023starcoder}). With the widespread adoption of these AI-powered code generators \cite{githubcopilot}, researchers focused more and more on investigating strategies to improve their predictions, both in terms of correctness (\ie the generated code is the one actually needed by the developer) and quality (\eg no security flaws introduced). Our work is mostly related to studies looking at the impact of the exploited training data on the correctness/quality of the generated predictions.

A key research direction focuses on improving prediction correctness by refining training data. Jain \etal \cite{jain2023llm} propose a data-cleaning pipeline that, by making the training code more structured and readable (\eg by renaming meaningless identifiers), leads to a higher percentage of correctly generated code. Xu \etal \cite{xu2023rethinking} develop pre-processing techniques to eliminate redundancy and enhance data diversity and instruction clarity, improving pass@1 on HumanEval \cite{chen2021evaluating}. Sun \etal \cite{sun2022importance} also present a (training) data cleaning framework focusing, however, on the code search task. They also report an increase in correct predictions as result of higher quality training data.
Song \etal \cite{song2024code} show that code comments in training data influence LLM-generated code accuracy and propose a data augmentation strategy to increase comment density. 

While related to our work, these studies do not look at the quality of the generated code in terms of non-functional requirements (\eg maintainability, security), but only at their functional correctness. 

Most studies on non-functional aspects of LLM-generated code focus on security risks, underlining the risky implications of integrating AI-generated code into real-world projects~\cite{pearce2022asleep, fu2023security, hajipour2024codelmsec, taeb2024assessing, majdinasab2023assessing}. Differently from our study, these works (i) are mostly observational studies, do not investigating how removing quality issues from training data affects the generated code; and (ii) are limited to security risks, dismissing other non-functional aspects. Three works partially addressed these limitations. 
Concerning the first limitation, Cotroneo \etal \cite{Cotroneo:icpc2024} explore the effects of training data poisoning on the security of code generated by DL models. They experiment with different amounts of vulnerable code in the fine-tuning data to assess to what extent AI models are prone to reproduce security flaws at inference time, finding a more than linear growth in the rate of vulnerable code between training and testing. 

Differently from \cite{Cotroneo:icpc2024}, our study does not artificially manipulate the training data by injecting security issues, but rather follows a classic training procedure reusing a large-scale code dataset available in the literature \cite{thestack}. Thus, while the work by Cotroneo \etal \cite{Cotroneo:icpc2024} is more focused on the possibility of succeeding in voluntarily poisoning DL models, our study aims at increasing awareness on the importance of curating training data rather than blindly reusing existing datasets. 

\eject

As for the second limitation, the works by Yeti{\c{s}}tiren \etal \cite{yeticstiren2023evaluating}, Siddiq \etal \cite{Siddiq:scam2022}, and {Liu \etal ~\cite{liu2024refining} extend the quality analysis of code generated by LLMs (GPT-Code-Clippy \cite{clippy}, Copilot \cite{githubcopilot}, ChatGPT \cite{chatgpt}) to additional non-functional aspects (on top of security), such as maintainability and code smells. Both studies, however, do not look at the relationship between the percentage of ``low-quality'' instances in the training sets and that in the code generated at inference time, as instead we do by training two different models: One on the whole dataset and one on a cleaned version of it.

Finally, worth mentioning are related studies looking at the \emph{memorization} phenomenon in LLM-based code generation \cite{yang2024unveiling,Ciniselli:msr2022}, meaning the tendency of the model to generate outputs that are near/exact replicas of examples seen during training. These works show that quite rarely LLMs tend to verbatim copy code from the training set \cite{Ciniselli:msr2022} and that removing duplicates from the training set can help in further minimizing such chances \cite{yang2024unveiling}.


\section{Conclusion and Future Work} \label{sec:conclusion}
We investigated the extent to which low-quality code seen at training time by DL-based code generators results in code quality issues generated at inference time. To do so, we collected 4.4M Python functions from a publicly available code dataset \cite{thestack}, showing that 4.98\% of its functions ($>$220k) are affected by at least one quality issue. We used this dataset to fine-tune DeepSeek-Coder \cite{guo2024deepseek} for the task of code generation, asking it to synthesize 551k Python functions. We found that 5.85\% of the functions generated by this model are affected by quality issues. We then removed from the above-mentioned dataset all functions affected by quality issues and fine-tuned a second DeepSeek-Coder on such a cleaned dataset. We showed that this model, when asked to generate the same 551k Python functions, outputs a significant lower number of low-quality functions, while not observing any drop in performance (\ie ability to generate functionally correct code).

Our findings have strong implications for tools (code generators) creators who, by curating the quality of their training data, can improve their tools' output quality without sacrificing functional correctness, leading to better code generation tools. 

Our future work mainly targets an increase in generalizability of our findings, experimenting with additional models and programming languages as well as with different types of quality issues. 


\section*{Acknowledgment}
This project has received funding from the European Research Council (ERC) under the European Union's Horizon 2020 research and innovation programme (grant agreement No. 851720), and from the \textit{IDA—Information Disorder Awareness} Project funded by the European Union-Next Generation EU within the SERICS Program through the MUR National Recovery and Resilience Plan under Grant PE00000014.

\eject

\bibliographystyle{IEEEtran}
\bibliography{biblio}

\end{document}